\documentclass{aastex631}

\begin{document}

\title{Oscillating dark energy in light of the latest observations and it's impact on the Hubble tension}

\author[0000-0002-3171-0369]{Mehdi Rezaei}
\affiliation{Iran Meteorological Organization\\
 Hamedan Research Center for Applied Meteorology\\
Hamedan, Iran}



\begin{abstract}

In this paper we have performed a comparative study of different types of oscillating dark energy models using the Metropolis algorithm of MCMC. Eight different oscillating parameterizations being examined herein which have demonstrated considerable ability to fit the overall cosmological observations including Pantheon sample of SnIa, Baryon Acoustic Oscillations, Cosmic Chronometer Hubble data and distance priors of Planck CMB. In order to compare the consistency of these models with observations, we have used both Akaike and Deviation information criteria. Although, the values of Akaike Information criteria for different models indicate that there is no support for oscillating DE models, Deviation Information criteria showed that there's significant support for some of these models. Our results showed that these models are capable of solving cosmic coincidence problem and alleviating the Hubble tension. Comparing the $H_0$ values obtained for different oscillating scenarios, with that of $\Lambda$CDM, we observed that our oscillating models led to $\bar{H_0}=69.78$ which is $0.29$ greater than $H_{0,\Lambda}$ and thus reduce the Hubble tension. Among all of models, Model(1), with $H_0=70.00\pm 0.71$ is the most capable of alleviating the $H_0$ tension. Forthermore, we examined our models assuming $H_0=73.0\pm 1.4$ from SHoES measurements. We find that adding this data point to our data combination, led to a $\Delta \bar{H_0}=0.95$ increase in $H_0$ value for different models.

\end{abstract}



\section{Introduction}
\label{sec:int}

Recent observations from the SHoES collaboration \citep{Riess:2021jrx} indicate that there is more than a $4-\sigma$ tension in the measurements of the Hubble parameter from Planck Cosmic Microwave
Background (CMB) data and direct local distance ladder measurements. We have $H_0=67.27 \pm 0.60$ km/s/Mpc from the $\Lambda$CDM based Planck measurements, while the results of SH0ES, using 70 long-period Cepheids in the Large Magellanic Cloud, give $H_0 = 74.03 \pm 1.42$ km/s/Mpc \citep{Planck:2018nkj,Planck:2018vyg,Riess:2019cxk}. 
Another reanalysis of the Planck data using the
Pearson correlation coefficient of T and E modes $R_l^{TE}$ leads to a new determination of Hubble constant as $H_0=67.5\pm 1.3$ km/s/Mpc \citep{LaPosta:2021gwp}

Besides these measurements, other independent measurements were done by the South Pole
Telescope (SPT-3G) and Atacama
Cosmology Telescope (ACT), two ground-based CMB experiments, have obtained $H_0=68.8 \pm 1.5$ and $H_0=68.4 \pm 1.5$ respectively \citep{SPT-3G:2021eoc,ACT:2020gnv}.
Furthermore, the other direct probes at lower redshift ranges have inferred lower values of $H_0$ which are in agreement with those predicted by $\Lambda$CDM based CMB data. One of these probes is based on a calibration of the Tip of the Red Giant Branch applied
to SNIa \citep{Freedman:2019jwv} which found $H_0= 69.8 \pm 0.8$ and the other is the latest strong lensing time delay data which found $H_0=67.4^{+4.1}_{-3.2}$ was reported in \cite{Birrer:2020tax}. On the other hand, the HoLiCOW collaboration, using strong gravitational lensing time delays, inferred $73.3^{+1.7}_{-1.8}$ km/s/Mpc as the value of the Hubble constant \citep{Wong:2019kwg}. Additionally, a measurement using the distance ladder method with
SNIa and Mira variable stars found $H_0 = 73.3 \pm 4.0$ km/s/Mpc \citep{Huang_2020}.
Given the importance of the $H_0$ tension, it is crucial, to investigate other cosmological models addressing this tension. These investigations are looking for the cosmologies that increase the $H_0$ value inferred from CMB data to resolve the tension\citep{Sola:2017znb,SolaPeracaula:2019zsl,SolaPeracaula:2021gxi,Sabla:2021nfy,Freese:2021rjq,Bargiacchi:2023jse,Briffa:2023ern}.
These attempts may be helpful in solving or alleviating the mentioned problem, as well as other problems that suffer the concordance model, namely $\sigma_8$ tension and the fine-tuning and cosmic coincidence problems. To this end, different cosmological models have been proposed. Many of these models have been studied in the literature including dynamical DE, running vacuum DE, different parameterizations for the equation of state (EoS) of DE, Holographic DE, Agegraphic DE, etc  
\citep{Perico:2013mna,Rezaei:2017yyj,Rezaei:2020mrj,Rezaei:2021qwd,Rezaei:2022bkb,Rezaei:2019hvb,Malekjani:2018qcz,Rezaei:2017hon,Moreno-Pulido:2022upl,Moreno-Pulido:2023ryo}.  One of the choices for solving the problems of $\Lambda$CDM, especially the coincidence problem is considering nonmonotonicity in the EoS of dark energy, same as an oscillating form for EoS of dark energy \citep{Linder:2005dw,Jain:2007fa,Pace:2011kb,Schmidt:2017iap,Tamayo:2019gqj}. Assuming an oscillating behavior for $w_{de}$, the accelerated expansion phase of the universe can be considered as one of the many
such phases occurring over cosmic history \citep{Rezaei:2019roe,Smith:2019ihp,Xu:2019amr}. This property of oscillating DE helps us to overcome the coincidence problem. Adding, results of \citep{Liu:2023ctw} using the latest Pantheon+ type Ia supernova sample and the Hubble parameter measurements indicate that a slightly oscillating EoS of dark energy around the $w_{de}=-1$ is favored. These motivate us to have a new look on oscillating DE models in the light of the latest observations and from a new point of view; those impact on the Hubble tension.
An oscillating Eos parameter was studied in \cite{Feng:2004ff} in which the EoS gets across $-1$. Their results show that this form of dark energy, dubbed oscillating quintom, can unify the early inflation and current acceleration of the universe, leading to the oscillations of the Hubble constant and a recurring universe.
Some oscillating scalar fields have been proposed in \cite{Dutta:2008px} as DE models. Authors of this work examine proposed models, with particular emphasis on the evolution of the ratio of the oscillation frequency to the expansion rate, and find that a subset of all three different classes of proposed oscillating models can be consistent with the observational data.
The possibility of detecting oscillating patterns in the EoS of the dark energy has been studied in \citep{Lazkoz:2010gz}. Their results show that data cannot discriminate between a cosmological constant and an oscillating equation of state.
A new dark energy model has been proposed in \cite{Tian:2019enx} for solving the cosmological fine-tuning and coincidence problems. The default assumption of this model is that the fine-tuning problem disappears if one does not interpret DE as vacuum energy and the coincidence problem will be solved if the Universe has several acceleration phases across the whole cosmic history. They study a quintessence model with approximately repeated double exponential potential, which only introduces one Planck scale parameter and three dimensionless parameters of order unity. Their results show that the oscillating model is able to explain the observed cosmic late-time acceleration.
Authors of \cite{10.1093/mnras/stz1229}, provide a general description for $w(z)$ in the form of a Fourier series. Their description generalizes some oscillating DE models and is in agreement with the $w(z)$ reconstructions. Using Markov chain Monte Carlo algorithm and different observational data they have compared the proposed model with $\Lambda$CDM, $w$CDM, and the standard Taylor approximation. They find that, even though there are extra parameters, there is a slight preference for the Fourier series compared with the $\Lambda$CDM model. This first analysis considering $w(z)$ as a Fourier series has its success in having a natural oscillatory behavior and being compatible with model-independent reconstructions\citep{10.1093/mnras/stz1229}.

In \cite{Tian:2020tur}, authors revisit the cosmological dynamics of an oscillating dark energy model and find that this model allows the Universe to evolve as an oscillating scaling solution (OSS) in the radiation era and as a chaotic accelerating solution (CAS) in the matter era. Mathematically, the transition from OSS to CAS is a route of period-doubling bifurcation to chaos. Upon their results, the global cosmological parameter constraints are practicable if the Universe evolves as an OSS in the radiation era and the late-time Universe described by a CAS can successfully explain the observed cosmic acceleration at low redshifts. The above oscillating DE model has been investigated in other studies and led to good results in predicting the tensor-to-scalar ratio 
$r$ and also as a new solution to the possible deuterium problem\citep{Tian:2021cqq,Tian:2022bwz}.

Following the above studies in this work, we wish to focus on Oscillating dark energy models as a proposal for solving the problems of $\Lambda$ cosmology, especially the $H_0$ tension. To this end, we have selected eight different oscillating DE parametrizations that have been proposed in the literature. Applying the MCMC algorithm we confront these DE models with the latest observations from different surveys. Adding to a comparative study between different models, we have examined the ability of this type of DE models to solve the $H_0$ tension. 

The paper is organized as follows: In Sec.\ref{sec:De}, we have reviewed eight different oscillating DE models as well as data sets we have used in our analysis. Sec.\ref{sec: results} gives the results of our analysis for different models under study and related discussions. Finally, we conclude our study in the Sec.\ref{sec:con}

\section{DE Models and Observations}
\label{sec:De}
In the first part of this section we describe different oscillating DE parameterizations. In the second part we introduce different data samples that we have used in our analysis.

\subsection{Oscillating DE parameterizations}
\label{sec:models}

Considering the universe including non-relativistic matter and dark energy in a spatially flat geometry, we have the Hubble parameter as follows:

\begin{equation}
    H^2=(\dot{a}/a)^2=\frac{8\pi G}{3}(\rho_m+\rho_{de})=\Omega_{m0}a^{-3}+\Omega_{de0}f(a).
	\label{eq:h}
\end{equation}

where $\rho_i$ is the energy density of fluid $i$, $\Omega_i=\frac{8 \pi G\rho_i}{3H^2}$ is the density parameter for fluid $i$ and $f(a)$ has the following form:
\begin{equation}
    f(a)=\exp{(-3 \int_{1}^{a} {\frac{1+w(a^{\prime})}{a^{\prime}}da^{\prime}})}.
	\label{eq:f}
\end{equation}
Determining EoS $w_d(a)$ for each of the DE scenarios and using the above equations, we can obtain the evolution of the Hubble parameter in the whole of cosmic history. Oscillating DE parametrizations which we have studied in this work are as follows:

\begin{itemize}

\item {\bf Model(1):} We have taken the second parametrization from \citep{Linder:2005dw} in which the authors examine the periodicity of the cosmic expansion in units of the e-folding scale $\ln{a}$ and select the following form for $w_{de}$:
\begin{equation}
    w_{de}(a)=w_0-A \sin{(B \ln{a})}.
	\label{eq:model1}
\end{equation}
where $w_0$ is the present value of $w_{de}$ and $A$ and $B$ are the amplitude and frequency of oscillation respectively.

\item {\bf Model(2):} We select the third parametrization from \citep{Ma:2011nc}.
This parameterization is a generalization of the CPL parametrization \citep{Chevallier:2000qy} in order to avoid the future unphysical divergence of this model.

\begin{equation}
    w_{de}(a)=w_0-A(a B \sin{(1/a)}).
	\label{eq:model2}
\end{equation}
where $w_0, A$ and $B$ are the present value of
$w_{de}(a)$, amplitude and frequency of oscillation respectively.
\item {\bf Model(3):} As the first parametrization in this work we consider a simple oscillating DE parametrization which was proposed by \citep{Feng:2004ff} with an EoS given by:

\begin{equation}
    w_{de}(a) = w_0 + w_1 \sin{(\frac{1}{a}-1)}.
	\label{eq:model3}
\end{equation}
The model has two free parameters $w_0$ and $w_1$, in which $w_0$ is the current value of EoS. At $a \rightarrow 1$ this model reduces to a linear parameterization while at $a<<1$ the
$w_{de}$ oscillates and will differ substantially
from the model with linear parameterization.

\item {\bf Model(4):} As the fourth parametrization in our study, we have investigated the following parametrization from \citep{Kurek:2008qt}:

\begin{equation}
    w_{de}(a)=-1+A \sin{(B \ln{(1/a)})}.
	\label{eq:m4}
\end{equation} 
The above parametrization in which oscillating behavior could not be
detected nowadays, has been compared with different linear and oscillating DE parametrizations in \citep{Kurek:2008qt} and obtained the best results between those DE scenarios.
\item {\bf Model(5):} The other parametrization we have studied in this work, is an oscillating
ansatz for $H(z)$ which is studied in \citep{Nesseris:2004wj}. Comparing this oscillating DE scenario with observations leads to the best results between 14 different DE models. 
In this scenario, $H(z)$ has the following form:

\begin{equation}
    H^2(a)=H_0^2[\Omega_{m0}a^{-3}+a_1 \cos{(a_2(1/a-1)^2+a_3)}+F_0],
	\label{eq:model5}
\end{equation} 
where $F_0=1-a_1 \cos{a_3}-\Omega_{m0}$.
This parametrization has $a_1, a_2$, and $a_3$ as the model parameters and those values will determine the behavior of the Hubble function at different epochs.

\item {\bf Model(6):} Same as Model (4), another parametrization has been investigated in \citep{Kurek:2008qt} as a purely oscillating dark energy with the following form:

\begin{equation}
    w_{de}(a)=w_0 \cos{(\alpha \ln{(1/a)})}.
	\label{eq:m6}
\end{equation} 
In this parametrization, we have two model parameters $w_0$ and $\alpha$ where the first one determines the value of EoS at the present time, $w_{de}(a=1)=w_0$.

\item {\bf Model(7):} Another oscillating parametrization which investigated in \citep{Feng:2004ff,Pan:2017zoh}, has two model parameters as following:

\begin{equation}
    w_{de}(a) = w_0 + \alpha(1-  \cos{(\ln{1/a})}).
	\label{eq:m7}
\end{equation} 

where $w_0$ is the current value of $w_{de}$ and $\alpha$ quantifies the dynamical character of the model. For $\alpha=0$ we acquire
$w_{da} = w_0$, while $\alpha \neq 0$ corresponds to a deviation from $\Lambda$CDM.

\item {\bf Model(8):} The last parametrization which proposed in \citep{Pan:2017zoh} is:

\begin{equation}
    w_{de}(a) = w_0 + \alpha (1-a) \cos{1/a}.
	\label{eq:m8}
\end{equation}

where $w_0$,the present value of EoS and $\alpha$ are the model parameters.
\end{itemize}
Assuming each of the above parametrizations, we will study the ability of oscillating DE scenarios to fit the latest observational data.

\subsection{Observational data}

In our analysis we have used the following data samples:

\begin{itemize}
\item {\bf SnIa:} We take the Pantheon sample of SnIa including 1048 data points within the redshift range $0.01 < z < 2.26$ from \cite{Scolnic:2017caz}.

\item {\bf Baryon acoustic oscillations (BAO):} Concerning this data sample we have used the radial component of the anisotropic BAOs have found from the measurements of the power spectrum and bispectrum from the BOSS Data Release 12 galaxy from \cite{Gil-Marin:2016wya}, the complete SDSS III Ly$\alpha$-quasar from \cite{duMasdesBourboux:2017mrl} and the SDSS-IV extended BOSS DR14 quasar sample from \cite{Gil-Marin:2018cgo}.

\item {\bf Cosmic Chronometers Hubble data(CCH):} This data sample was obtained from the spectroscopic techniques applied to passively–evolving galaxies, i.e., galaxies with old stellar populations and low star formation rates. In this work we have used CCH sample including 37 data points from the following references: \cite{Jimenez:2003iv,Simon:2004tf,Stern:2009ep,Moresco:2012jh,Zhang:2012mp,Moresco:2015cya,Moresco:2016mzx,Ratsimbazafy:2017vga}.

\item {\bf $H_0$ from SHoES measurements:} In order to study the effects of SHoES measurements on our analysis, we have used $H_0=73.0 \pm 1.4$ from \cite{Riess:2020fzl}.

\item {\bf Cosmic microwave background(CMB):} We have used the CMB distance prior from the final Planck 2018 release instead of full CMB data. The distance priors that are derived from the base $\Lambda$CDM model can be used to replace the global fitting of full data released by Planck in 2018 for the other DE models \citep{Chen:2018dbv}.

\end{itemize}

\section{results and discussions}
\label{sec: results}

Here we present the results of our analysis in two different parts. In the first part, we report our numerical results and compare different DE scenarios under study in fitting observations. In the second part, we compare the ability of oscillating DE models to solve cosmological problems, especially the Hubble tension.

\subsection{Oscillating DE and observations}
\begin{figure*}
	\includegraphics[width=5.5cm]{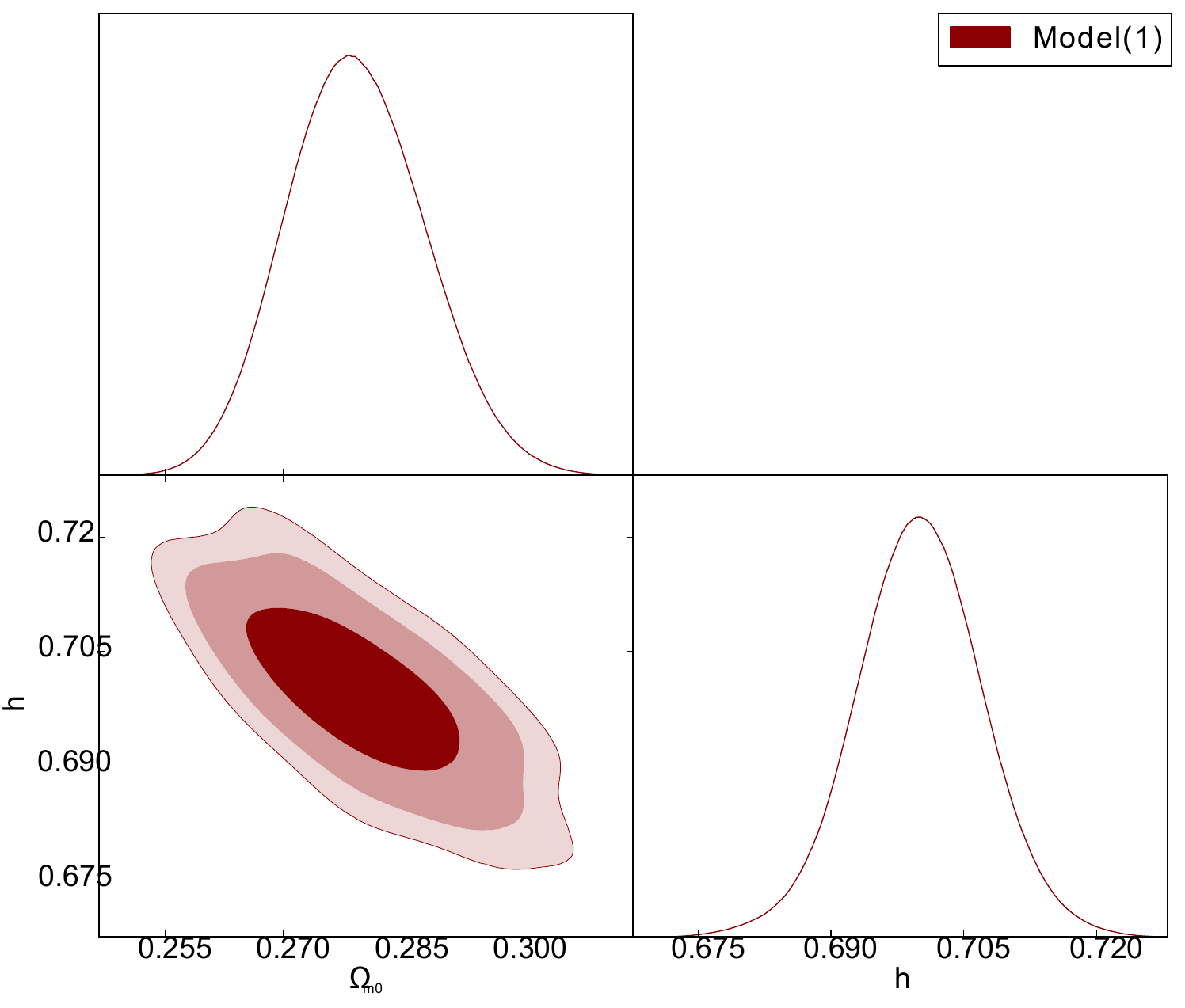}
    \includegraphics[width=5.5cm]{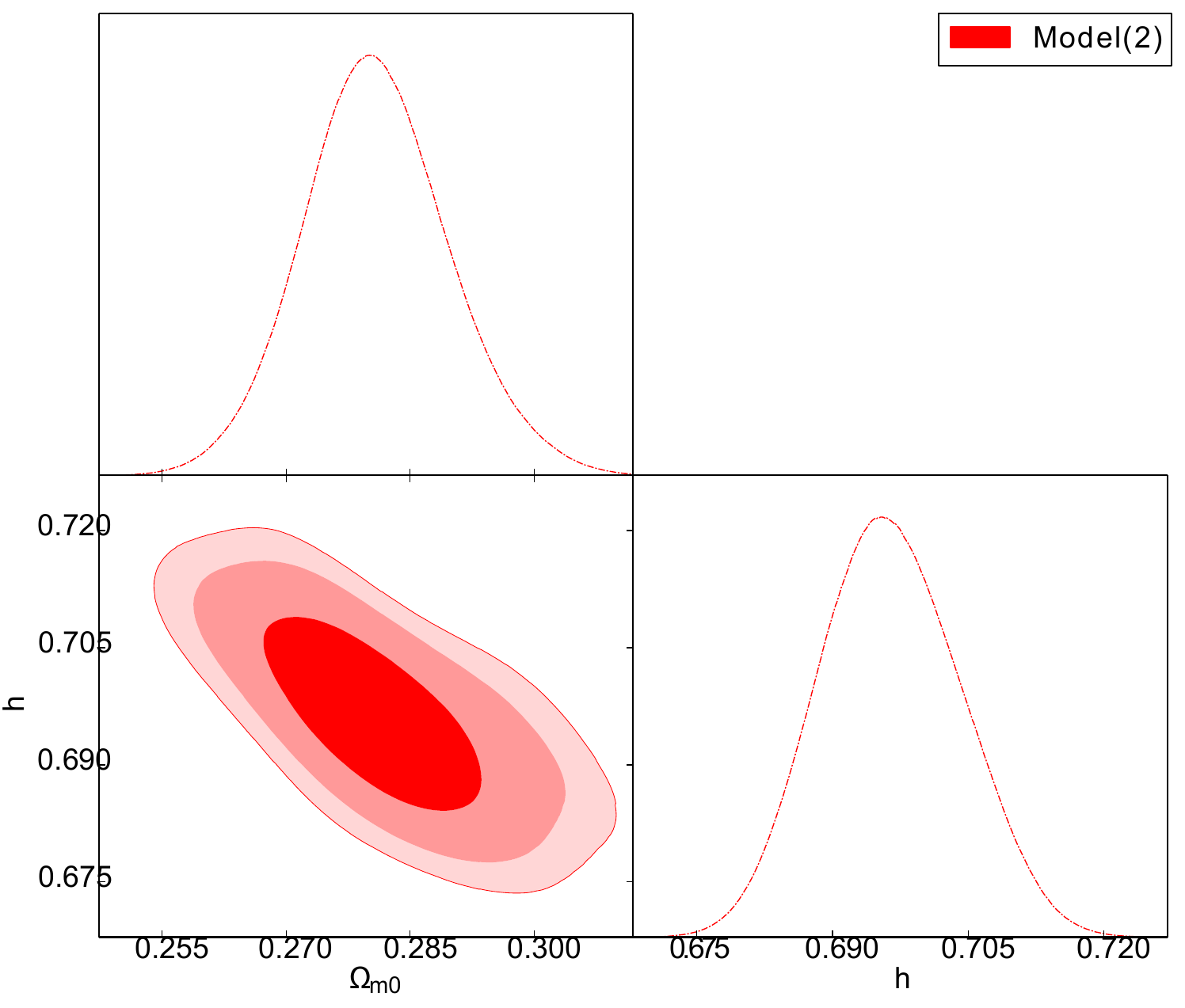}
	\includegraphics[width=5.5cm]{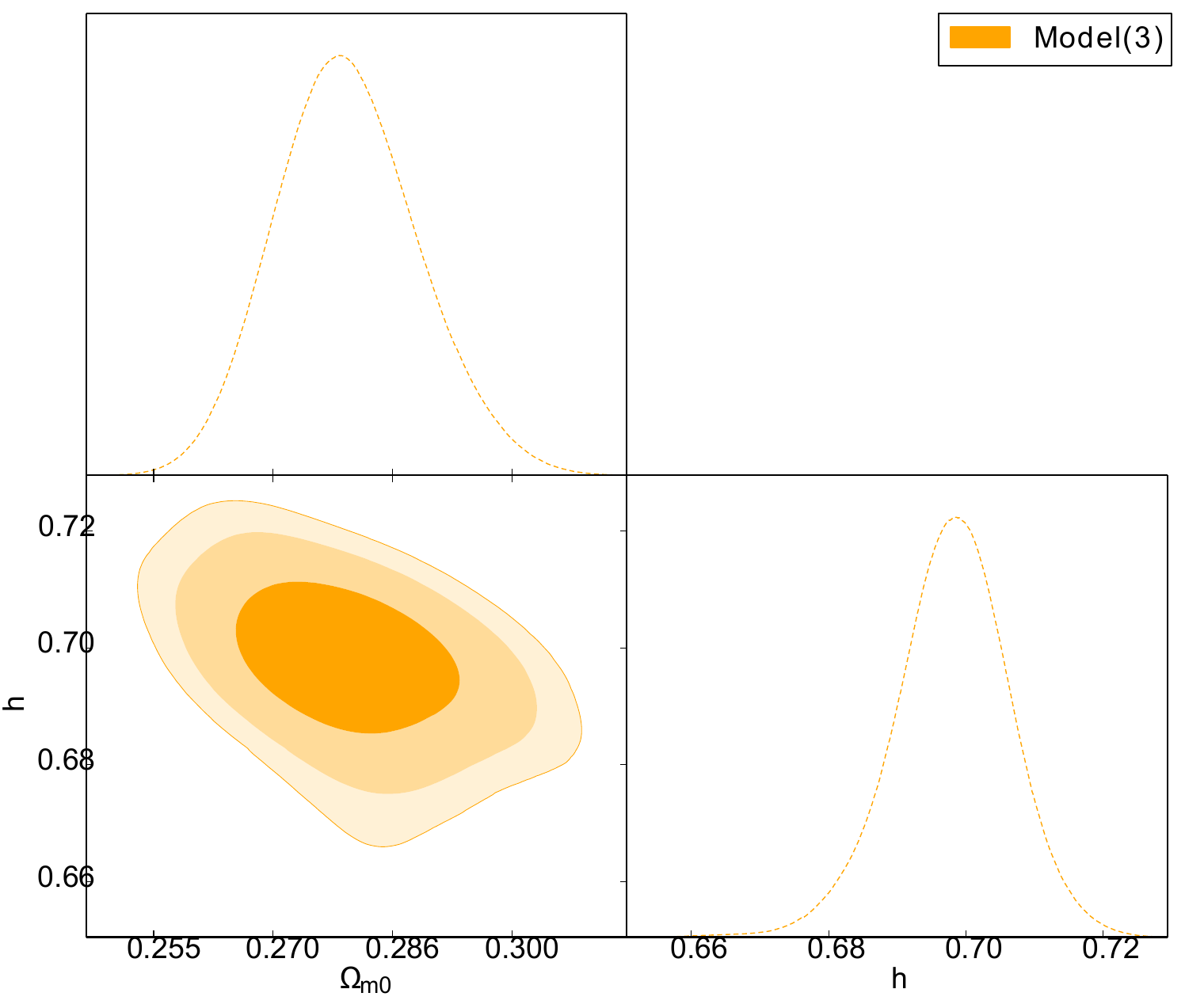}
	\includegraphics[width=5.5cm]{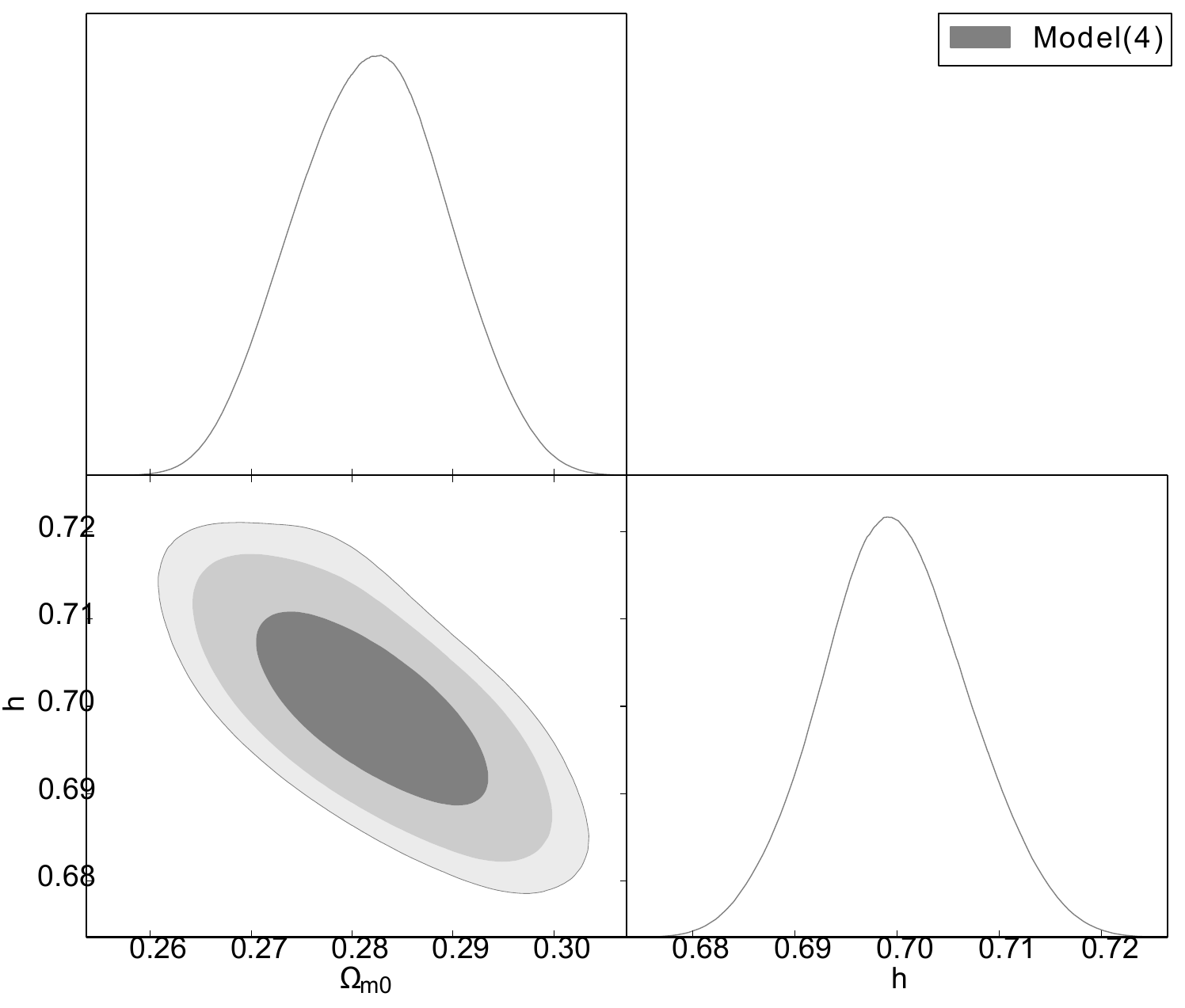}
	\includegraphics[width=5.5cm]{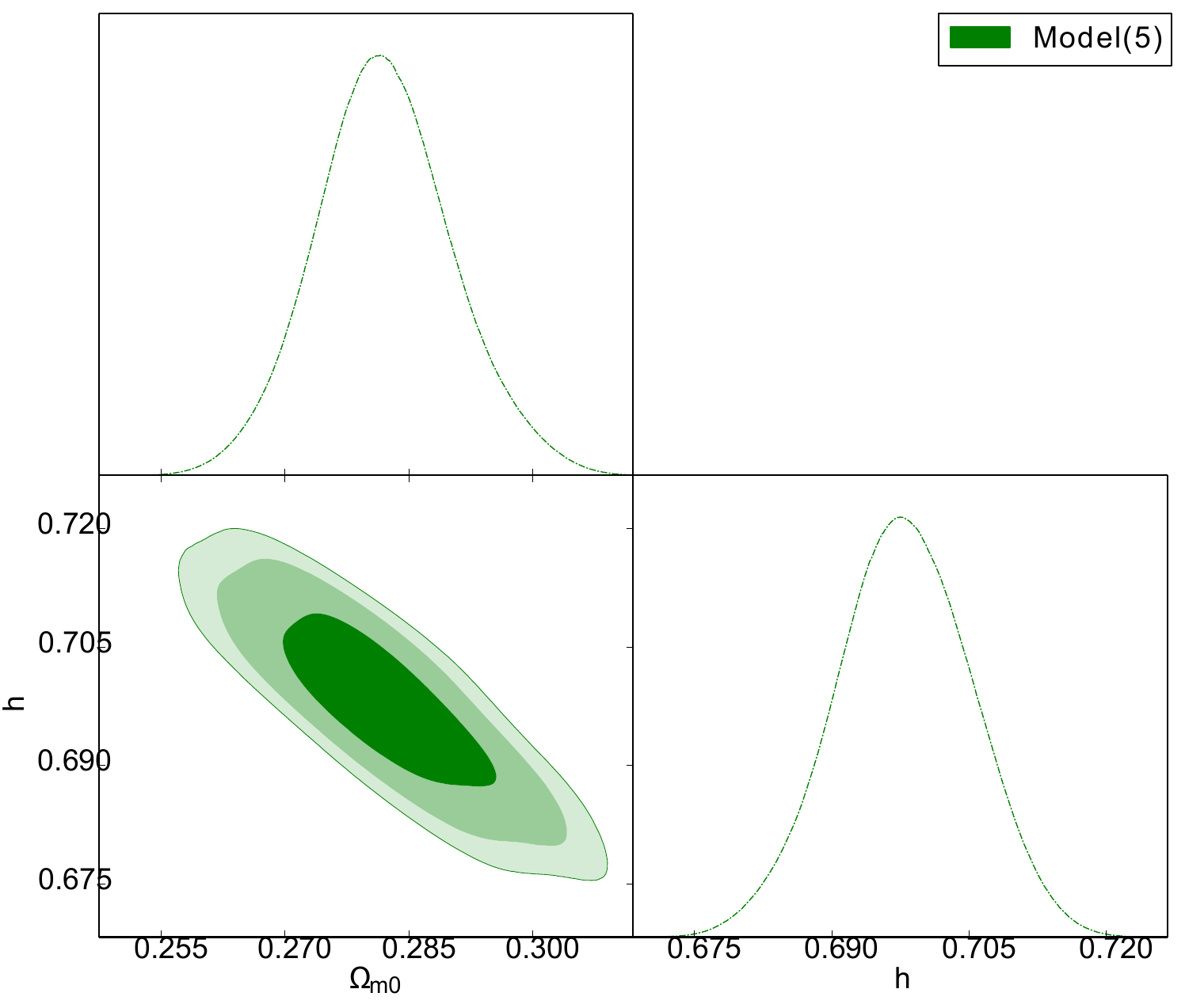}
	\includegraphics[width=5.5cm]{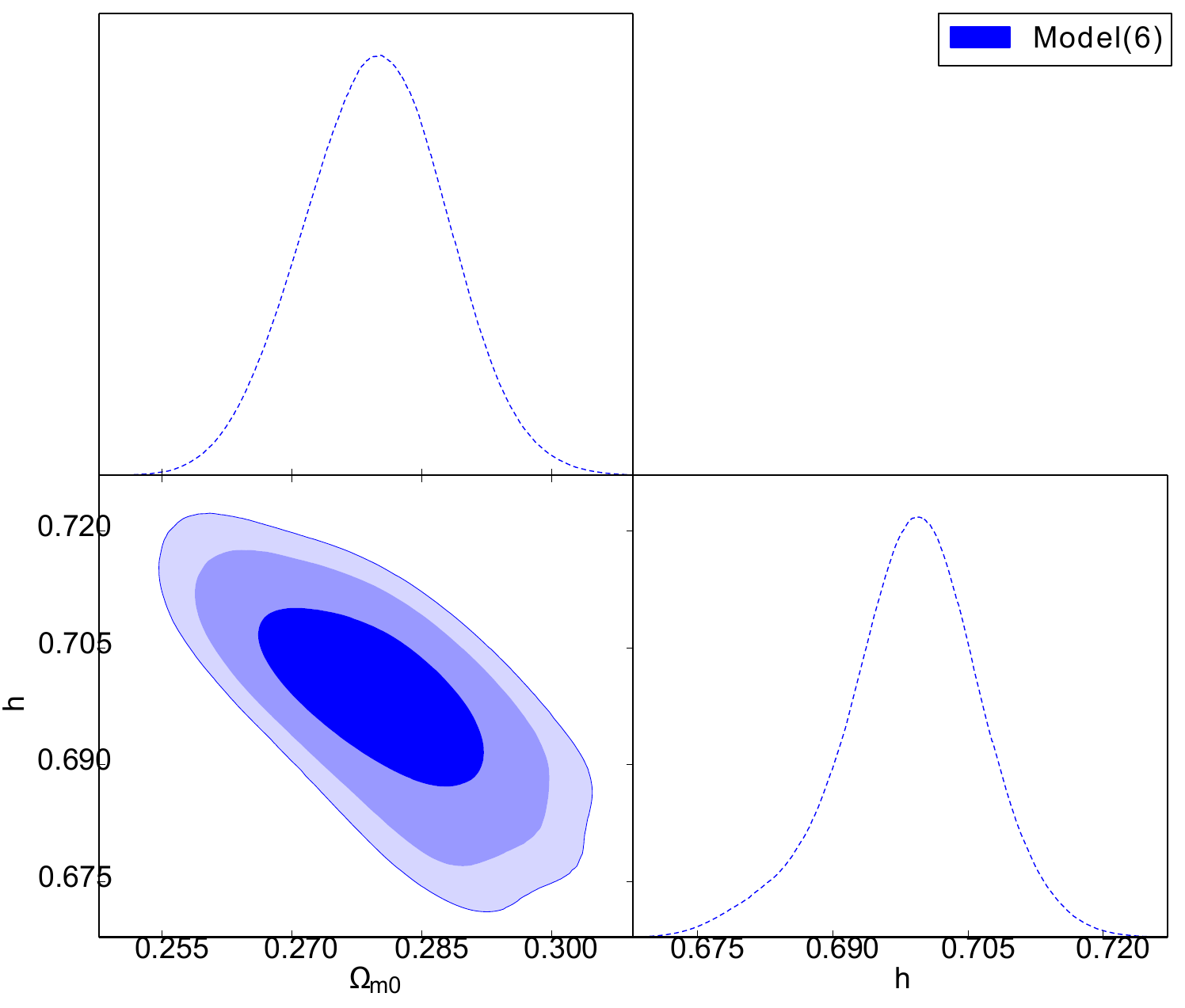}
	\includegraphics[width=6cm]{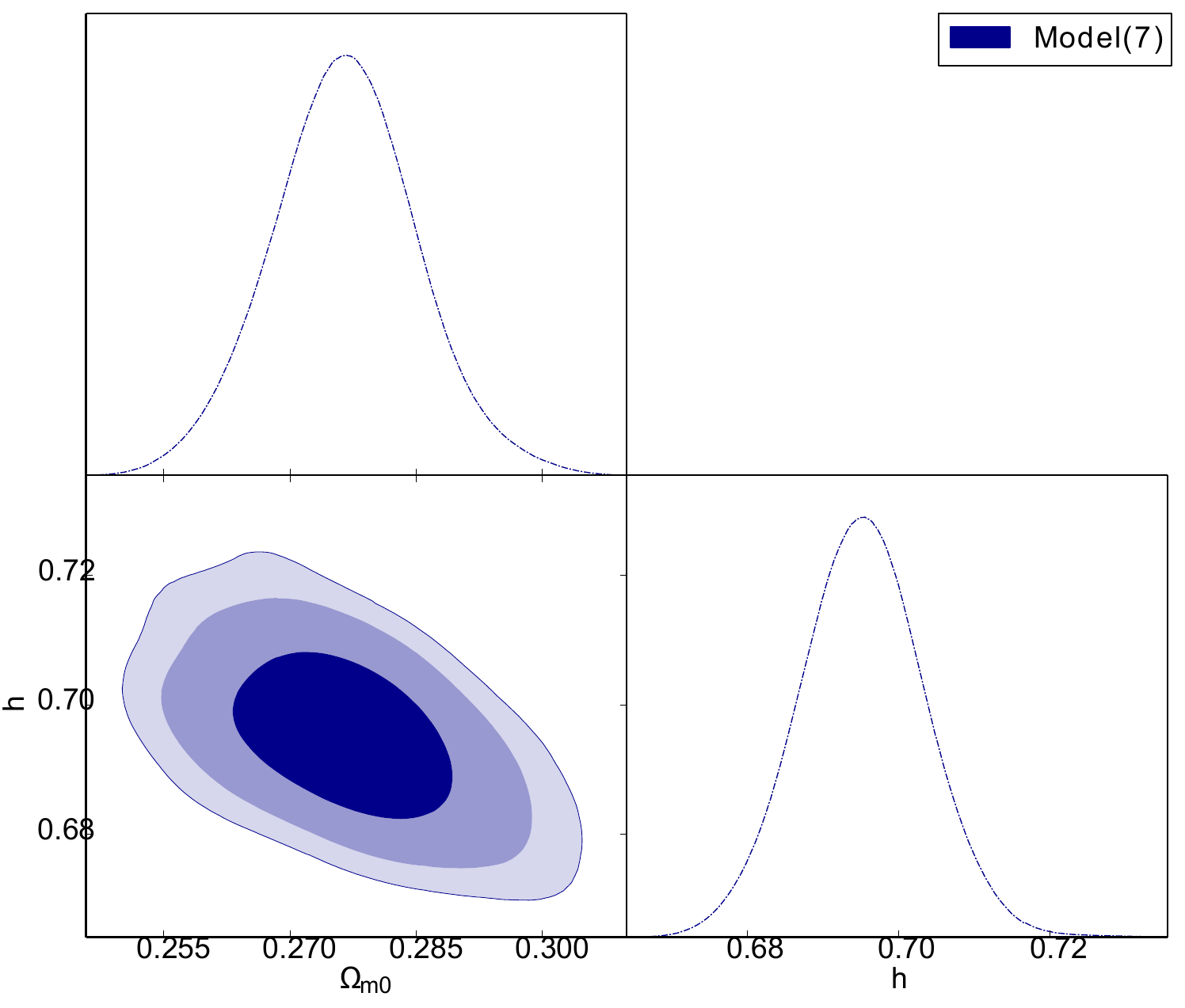}
	\includegraphics[width=6cm]{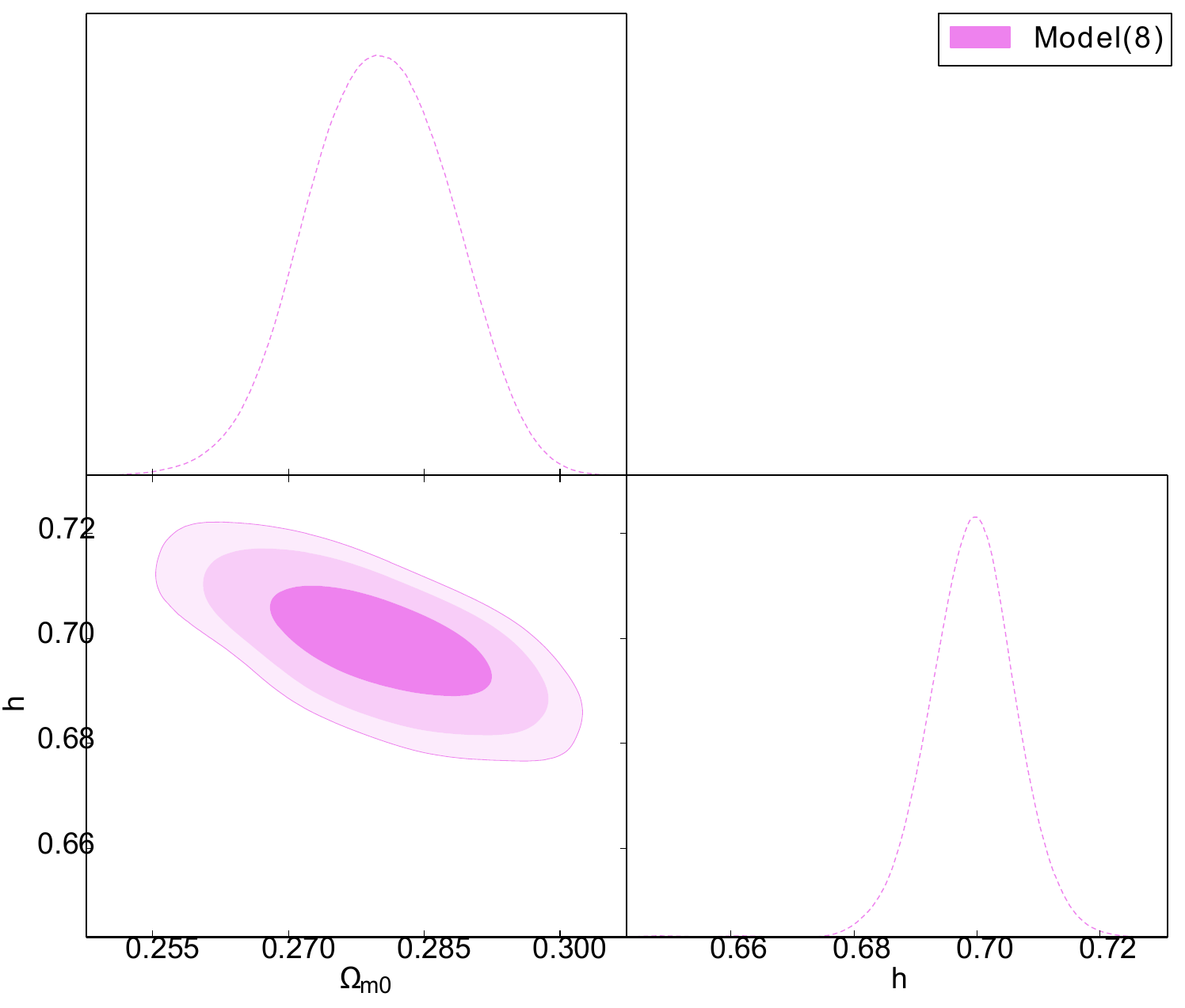}
	\includegraphics[width=6cm]{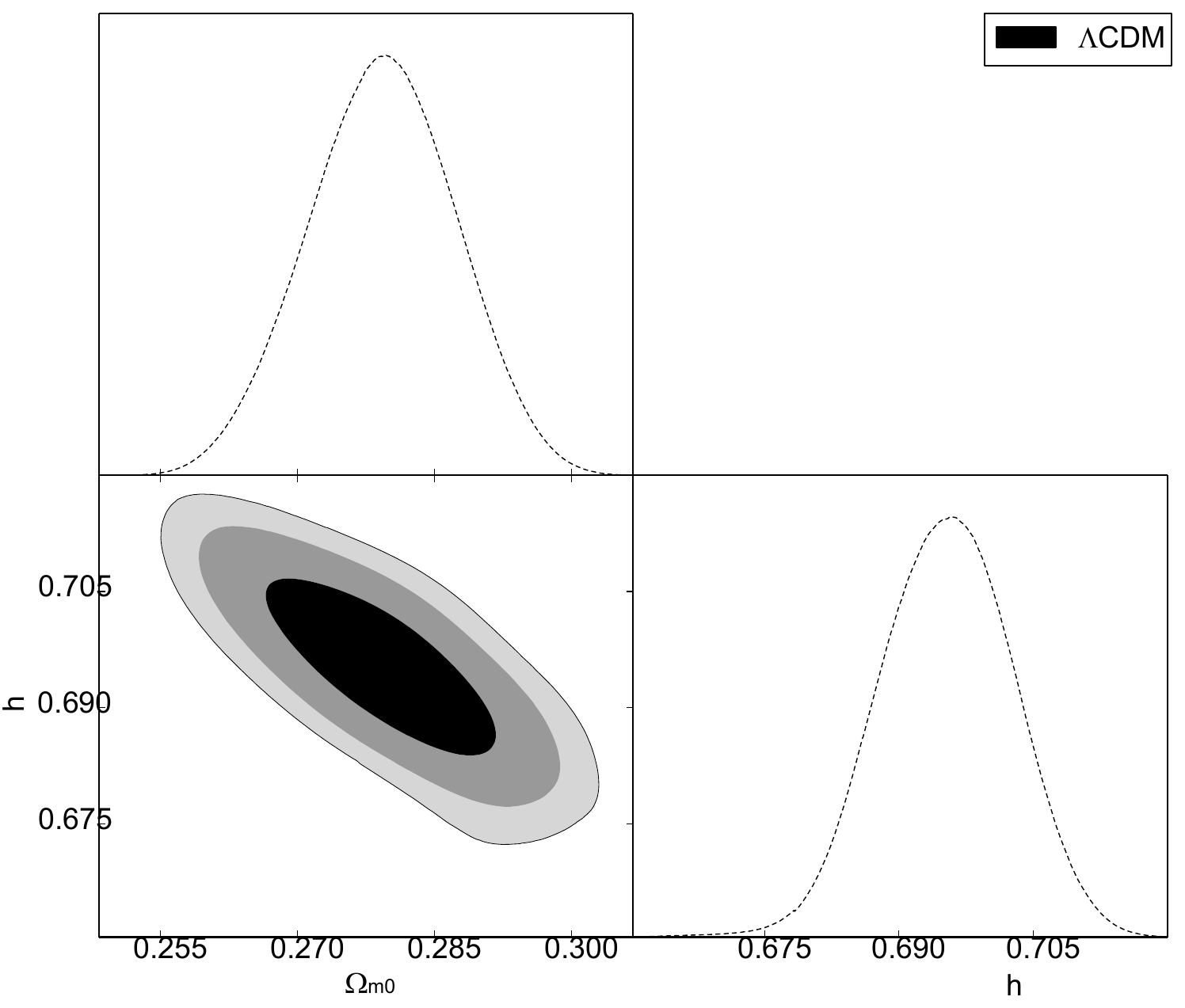}
 \caption{Contour plots for the confidence levels of the model parameters $\Omega_{m0}$ and $h$ up to $3\sigma$, obtained for different oscillating DE scenarios investigated in this work.}
    \label{fig:contour}
\end{figure*}

In order to put constraints on the model parameter we have used the Metropolis algorithm of MCMC \citep{10.1063/1.1699114}. As the data set, we have used a data combination including 1094 data points of SN+BAO+CCH+CMB. To search all the parameter space and prevent the MCMC from selecting a local best fit, in our algorithm we did not set any limitations on the priors. In each of the analyses, we have 150000 chains which provides sufficient iterations in order to obtain the best fit parameters.
Traceplot corresponding to a Markov chain, provides a visual way to inspect sampling behavior and convergence. It is one intuitive and easily implemented diagnostic tool that plots the parameter value against the iteration number. When the traceplot moves around the mode of distribution, we find that our model has converged. It is easy to see in Figs.\ref{fig:contour} \& \ref{fig:example_figure} that in our analysis we obtained traceplots with normal distribution which means our models have converged.
In order to examine the effect of choosing initial values in MCMC, we have repeated our algorithm with different initial values for the $\Lambda$CDM model. Our results indicate that changing initial values may lead to a change in the best fits up to $1 \%$. Although choosing convenient initial values led to saving time in reaching the convergence, it hasn't significant effect on the numerical results. The best value of model parameters and their 68\% CL constraints are reported in Tab.\ref{tab:best}. In this table, we only presented those parameters which are common between all models. In different plots of Fig.\ref{fig:contour}, we showed One-dimensional marginalized posterior distributions of $\Omega_{m0}$ and $h$, and the two-dimensional joint contours between these model parameters ($\Omega_{m0}$ and $h$) at $68\%, 95\%$ and $99.7\%$ CL of DE models.    
There are several methods for selecting the best model among different probable models. The simplest way is to compare the $\chi^2$ values obtained from MCMC chains. In this way, the model with the smallest value of $\chi^2$ is selected as the best one. But in this manner, we did not assume the extra model parameters. In order to assume the effect of extra model parameters, different information criteria have been proposed in literature;  Akaike Information Criteria (AIC) \citep{Akaike:1974}, Corrected Akaike Information Criteria ($AIC_c$) \citep{CAVANA:1997}, Bayes Information Criteria (BIC) \citep{Schwarz:1974}, Deviance Information Criteria (DIC) \citep{spiegelhalter2002bayesian}, etc [see detailed discussions in Refs. \cite{Rezaei:2021qpq}]. Among these criteria in this work, we have used the most popular information criterion AIC and DIC which is understood as a Bayesian version of AIC. Like AIC, it trades off a measure of model adequacy against a measure of complexity and is concerned with how hypothetically replicated data predict the observed data \citep{LI2020450}. DIC is an asymptotically unbiased estimator of the expected Kullback-Leibler divergence between the data-generating process and the plug-in predictive distribution.

The AIC is defined as: 
\begin{eqnarray}
{\rm AIC} = \chi^2_{\rm min}+2K\;.
\end{eqnarray}
in which $K$ is the number of model parameters. Assuming this information criterion, the preferred model is the one that has the minimum value of AIC. 
The other criterion, DIC is defined as \citep{Liddle:2007fy}
\begin{eqnarray}
{\rm DIC} = D(\bar{{\bf p}})+2C_{\rm B}\;.
\end{eqnarray}
where $C_{\rm B}=\overline{{D({\bf p})}}-D(\bar{{\bf p}})$ 
is the Bayesian complexity and over-lines imply the standard mean value. 
The Bayesian deviation, $D({\bf p})$ is expressed as $D({\bf p})=\chi^2_{\rm t}({\bf p})$ in the case of an exponential class of distributions \citep{Rezaei:2021qpq}. We have reported the AIC and DIC values for all of the models under study. In order to have a comparison between different models, we calculated $\Delta$AIC ($=$ AIC (model) $-$ AIC ($\Lambda$CDM)) and $\Delta$DIC ($=$ DIC (model) $-$ DIC ($\Lambda$CDM)).
These resulting differences are used to determine the ``level of support for'' each model with respect to the concordance model [see more details in \citep{Rezaei:2019xwo}].

\begin{table*}
\centering
\caption{Summary of the 68\% CL constraints on various free parameters of the Oscillating DE scenarios and $\Lambda$CDM for the combined dataset SN+BAO+CCH+CMB. We note that all of $\Delta$ICs were calculated with respect to $IC_{\Lambda}$.  }
 \begin{tabular}{c  c  c c c c c c c}
 \hline

\multicolumn{9}{c}{SN+BAO+CCH+CMB} \\
 \hline 
 Model & $\Omega_{m,0}$  & $h$  &   & $\chi^2_{min}$ 
 & $AIC$ & $DIC$ & $\Delta AIC$ & $\Delta DIC$\\
 \hline 
 Model(1) & $0.2792\pm 0.0086$ & $0.7000\pm 0.0071$ &  & $1062.2$ & $1074.2$  & $1075.0$& $6.8$& $2.0$ \\
 
 Model(2) & $0.2808\pm 0.0086$ & $0.6947 \pm 0.0077$ &  & $1063.6$ & $1075.6$  & $1075.4$& $8.2$& $2.4$ \\

 Model(3) & $0.2792^{+0.0083}_{-0.0095}$ & $0.6980^{+0.0086}_{-0.0077}$ &  & $1063.6$ & $1073.6$  & $1078.8$& $6.2$& $5.8$ \\

 Model(4) & $0.2819\pm 0.0073$ & $0.6998\pm 0.0071$ &  & $1063.8$ & $1073.8$  & $1074.2$& $6.4$& $1.2$ \\

 Model(5) & $0.2820\pm 0.0083$ & $0.6979^{+0.0074}_{-0.0067}$ &  & $1064.1$ & $1076.1$  & $1072.3$& $8.7$& $-0.7$ \\

 Model(6) & $0.2797\pm 0.0080$ & $0.6987^{+0.0081}_{-0.0067}$ &  & $1064.4$ & $1074.4$  & $1074.4$& $7.0$& $1.4$ \\

 Model(7) & $0.2765\pm 0.0086$ & $0.6942\pm 0.0082$ &  & $1062.4$ & $1072.4$  & $1076.0$& $5.0$& $3.0$ \\

Model(8) & $0.2801\pm 0.0078$ & $0.6991\pm 0.0074$ &  & $1064.1$ & $1074.1$  & $1075.3$& $6.7$& $2.3$ \\
  
$\Lambda$CDM & $0.277\pm 0.0084$ & $0.6949\pm 0.0082$ &  & $1061.4$ &  $1067.4$ & $1073.0$& $0.0$& $0.0$ \\
\hline 
\hline 
\end{tabular}\label{tab:best}
\end{table*}
As one can see in Tab.\ref{tab:best}, $\Lambda$CDM with $AIC=1067.4$ is the best model, while for all of the oscillating models $\Delta$AIC$ >5$. This means that from AIC point of view, there is "Considerably less support for" these models compared with the concordance model. This result was predictable because our oscillating DE models have $2-3$ more extra parameters, which led to an increment in the AIC value. 
On the other hand upon DIC results, the performance of oscillating models is good. In this case, Model (5) with $\Delta$DIC$=-0.5$ is the best model followed by $\Lambda$CDM. Also, we have found "Significant support" for models (1), (4), and (6) because of $\Delta$DIC$ <2$. For models (2), (7) and (8) we have "Less support" and finally we have "Considerably less support" for Model (3).

\subsection{Oscillating DE and cosmological problems}
\begin{figure*}
\includegraphics[width=9cm]{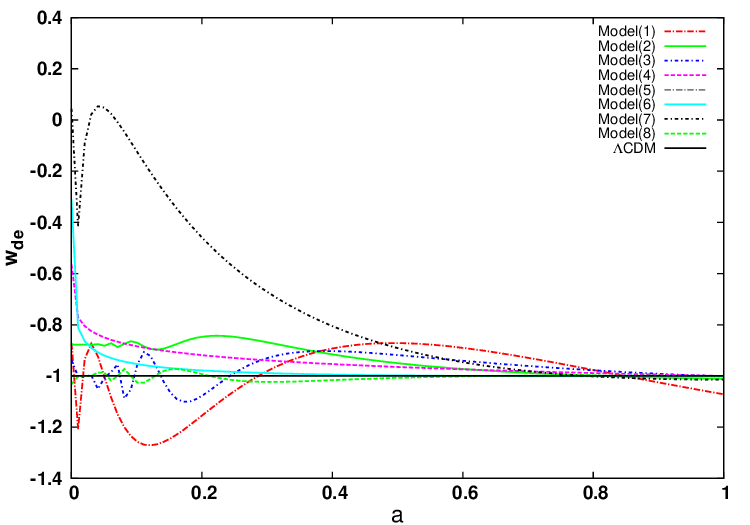}
    \caption{The evolution of EoS parameter $w_{de}$ obtained for different Oscillating DE models as well as $\Lambda$CDM upon their best fit parameters.}
    \label{fig:w}
\end{figure*}

\begin{figure*}
\includegraphics[width=10cm]{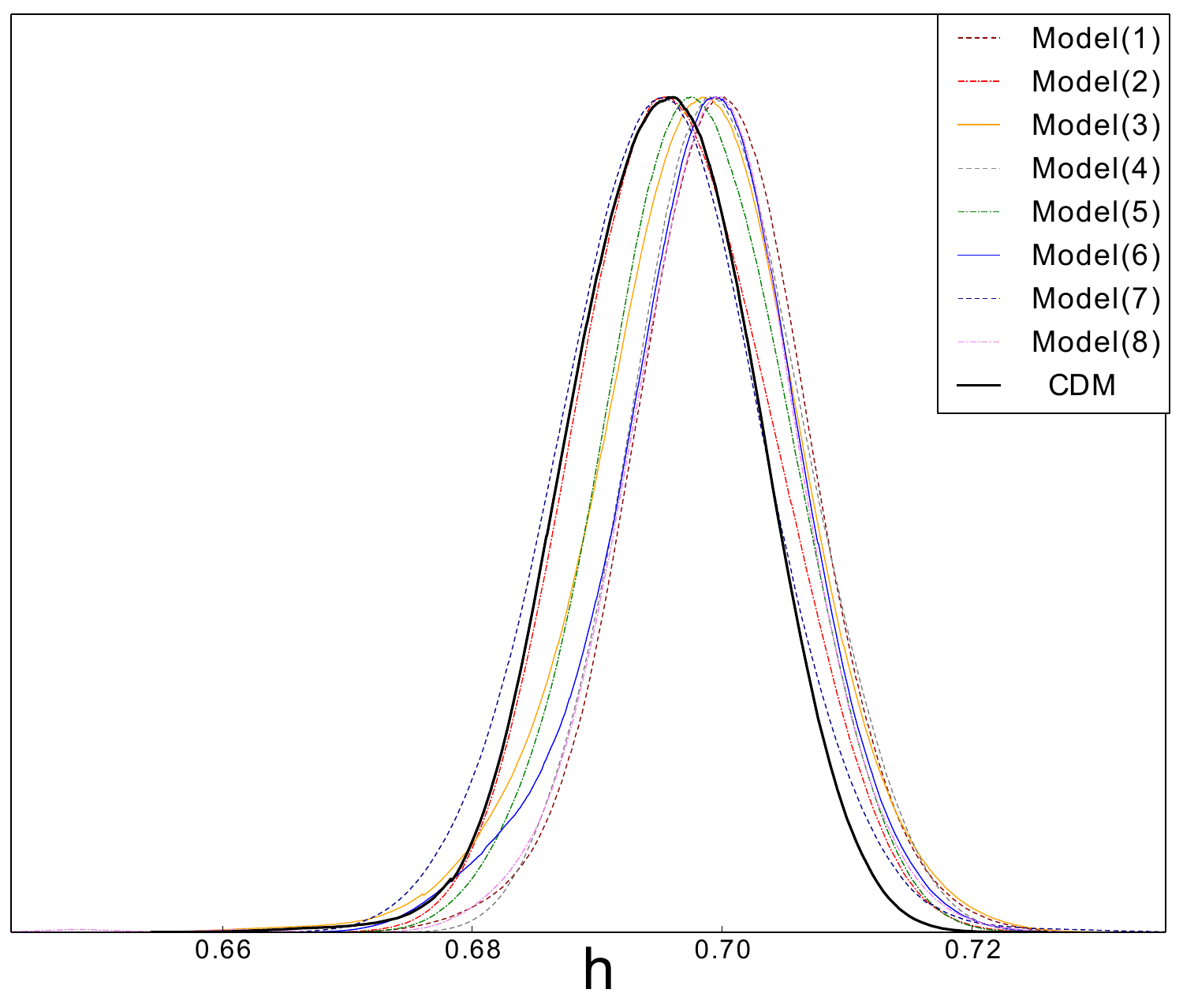}
    \caption{One-dimensional marginalized posterior distributions of the $h$ obtained for different Oscillating DE models as well as $\Lambda$CDM.}
    \label{fig:example_figure}
\end{figure*}

\begin{figure}
\includegraphics[width=9cm]{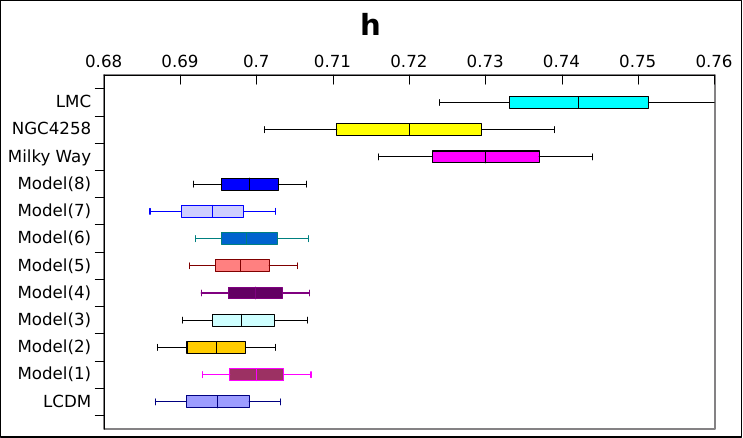}
\includegraphics[width=9cm]{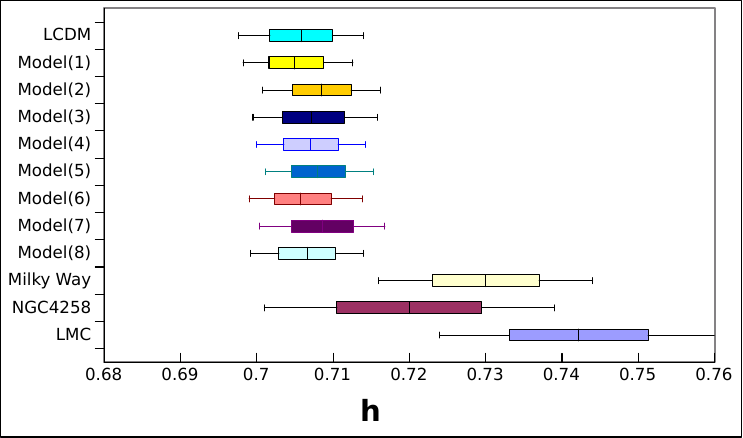}
    \caption{The values of $h(=H_0/100)$, obtained for $\Lambda$CDM and Oscillating models with their $1-\sigma$ uncertainties. The results have compared with those obtained from the SHoES measurements, based on the Cepheid calibrations for the LMC, NGC 4258, and the Milky Way. The upper panel shows the results which obtained using SN+BAO+CCH+CMB, while the lower panel shows the results were obtained using SN+BAO+CCH+CMB+$H_0$.}
    \label{fig:shoe}
\end{figure}

The considerable fact that the energy density of DE and that of dark matter are of the same order around the present time, is called the “Cosmological Coincidence Problem" (CCP). As a simple result, one can conclude from this fact, that we are living in a very special time of cosmic history. In $\Lambda$ cosmology where the DE density $\Lambda$ is constant while the density of dark matter scales with the $a^{-3}$, this appears to be a coincidence \citep{Velten:2014nra}. Because it requires extreme fine-tuning in the initial conditions of the early Universe. This point becomes more interesting knowing that in the very early and in the far-future Universe, both of the energy densities differ by many orders of magnitude. The $\Lambda$ cosmology offers no explanation for the peculiar circumstance that $\Omega_{\Lambda} \approx \Omega_m$ at the present time.
On the other hand, oscillating DE scenarios offer a new response to the CCP. Upon these scenarios, the current state of the Universe, with $\Omega_{\Lambda} \approx \Omega_m$, has happened previously and will happen again. Therefore, no fine-tuning is required, so the CCP will resolve in DE scenarios with oscillating EoS parameter. In Fig.\ref{fig:w} we showed the evolution of $w_{de}$ for models under study. Although at present time $(a \approx 1)$, $w_{de}$ for all
of oscillating models, come close to $w_{\Lambda}=-1$, but at smaller scale factor, oscillating model deviate from concordance models. This deviation from $w_{\Lambda}=-1$
changes behavior of $\Omega_{de}$ in these models which finally led to solving CCP.
As we mentioned in section \ref{sec:int}, the observed tension in the value of the Hubble parameter obtained from Planck measurements and that of Cepheid-calibrated SnIa distances, has reached a statistical significance
of $5-\sigma$. We note that the Planck measurement was done upon the $\Lambda$
cosmology, while the SH0ES measurement does not. This situation motivated the science community to develop new
extensions of the $\Lambda$ cosmology attempting
to resolve or alleviate this tension by increasing the $H_0$ value. As one of the main goals of this work, we have examined the ability of oscillating DE models to resolve $H_0$ tension. In this way, we selected $h(=\frac{H_0}{100})$ as one of the free parameters of our models and then we put constraints on it using observations. In Fig.\ref{fig:example_figure} we showed One-dimensional marginalized posterior distributions of the $h$ for different DE models. It is easy to see that for all of the oscillating models except Model (2) and Model (7), the best value of $h$ is greater than $0.6949$ obtained for $\Lambda$CDM.
In order to have a quantifying view on the ability of models to alleviate Hubble tension, we will compare the results of our models with those of $\Lambda$ based Planck measurements and SHoES measurements from \cite{Riess:2020fzl}. While there is a 4.1-$\sigma$ tension between the results of Planck and SHoES, we obtained less values of tensions upon oscillating models. As one can see in Tab. \ref{tab:best}, the best value of $H_0(=100h)$ for all the oscillating models is greater than 67.27 which obtained by Planck measurements. In Model (1), as the more capable model in solving Hubble tension, we have $H_0=70.00 \pm 0.71$ that indicate a $2.14-\sigma$ tension with SHoES results. After Model (1), we have Model (4) which decreases the Hubble tension to 2.16-$\sigma$ with respect to SHoES result. Model (8), Model (6), Model (3), Model (5) and Model (2) stand in the lower position in our ranking as the cosmological models which can alleviate the Hubble tension. Finally, we have Model (7) as the worst model which decreases the Hubble tension to 2.56-$\sigma$.
From another point of view, we can compare the ability of oscillating models in alleviating Hubble tension with $\Lambda$CDM. As we observe in Tab.\ref{tab:best}, in our analysis we obtained $H_0=69.49 \pm 0.82$ for concordance model which shows a 2.51-$\sigma$ tension with respect to SHoES result. So, we can claim that oscillating DE scenario in 6 cases lead to better results compared to concordance model. In the other meaning, just for Model (2) and Model (7) we observed tensions greater than 2.51-$\sigma$, while for other oscillating models Hubble tension falls below 2.51-$\sigma$.
Summarizing all of the above, we found that oscillating DE models reduces the observed tension between $\Lambda$ based Planck measurements and SHoES, from 4.1-$\sigma$ to less than 2.56-$\sigma$. In comparison with $\Lambda$CDM, all of oscillating models except Model (2) \& (7), lead to smaller value of Hubble tension between 2.14 to 2.29-$\sigma$. Therefore, Oscillating DE models can reduce the Hubble tension.  
In order to have a comparison with SHoES measurements, we have plotted the $H_0$ values obtained for studied models and their uncertainties, with those obtained from the SHoES measurements, based on the Cepheid calibrations for the LMC, NGC 4258, and the Milky Way. From these measurements, we have $H_0=74.22 \pm 1.82$ \citep{Riess:2019cxk}, $H_0=72.0 \pm 1.9$ \citep{Riess:2019cxk}, and $H_0=73.0 \pm 1.4$ \citep{Riess:2020fzl} for the LMC, NGC 4258, and the Milky Way respectively. The results can be seen in the upper panel of Fig.\ref{fig:shoe}. We observe that assuming obtained $H_0$ values and their $1-\sigma$ uncertainty, all of the oscillating DE models investigated in this work, can reduce the $H_0$ tension. These results indicate that oscillating DE scenario can be an alternative candidate for alleviating Hubble tension in cosmology.
In order to see the effect of including SHoES measurements in our data sets, we added $H_0=73.0 \pm 1.4$ from \citep{Riess:2020fzl} to our data sets. We reported the results of this part of the analysis in Tab.\ref{tab:besth}. In this table, we compared the best fit values of $h$ obtained using SN+BAO+CCH+CMB and those obtained using SN+BAO+CCH+CMB+$H_0$. Our results indicate that adding SHoE data to the data samples leads to an increase in the best fit of parameter $h$. This increment varies between $\Delta h=0.005$ to $\Delta h=0.0144$ for different models which means a $0.5$ to $1.44 [km/s/Mpc]$ increase in $H_0$ value.
The maximum change ($\Delta h$)occurred in Model (7), while the minimum change occurred in Model (1). In the case of $\Lambda$CDM we observed a $0.0109$ increase in the best value of $h$ after adding $H_0$ from SHoE to our data. In the lower panel of Fig.\ref{fig:shoe}, we compared our results for different DE models after assuming SHoES in our data sets. 
Differences between the upper panel (data samples without SHoES) and the lower panel (data samples with SHoES) show the effect of including $H_0$ in our data sets clearly.
\begin{table}
\centering
\caption{Summary of the 68\% CL constraints on parameters $h$ of the Oscillating DE scenarios and $\Lambda$CDM for two different combined datasets, SN+BAO+CCH+CMB (data sample I) and SN+BAO+CCH+CMB+$H_0$ (data sample II). Note that $\Delta h = h_{data sample II}-h_{data sample I}$.  }
 \begin{tabular}{c  c  c c c}
 \hline

Model & data sample I &  & data sample II & $\Delta h$ \\
 \hline 
 Model(1) & $0.7000\pm 0.0071$ &  & $0.7050^{+0.0076}_{-0.0068}$ & $0.0050$ \\
 
 Model(2)  & $0.6947 \pm 0.0077$ &  & $0.7085^{+0.0077}_{-0.0069}$& $0.0138$ \\

 Model(3) & $0.6980^{+0.0086}_{-0.0077}$ &  & $0.7072^{+0.0081}_{-0.0078}$& $0.0092$  \\

 Model(4) & $0.6998\pm 0.0071$ &  & $0.7071\pm 0.0072$& $0.0073$  \\

 Model(5) & $0.6979^{+0.0074}_{-0.0067}$ &  & $0.7079 \pm 0.0071$& $0.0100$  \\

 Model(6) & $0.6987^{+0.0081}_{-0.0067}$ &  & $0.7057^{+0.0079}_{0.0064}$ & $0.0070$ \\

 Model(7) & $0.6942\pm 0.0082$ &  & $0.7086 \pm 0.0065$ & $0.0144$ \\

Model(8) & $0.6991\pm 0.0074$ &  & $0.7066^{+0.0086}_{0.0075}$ & $0.0075$ \\
  
$\Lambda$CDM & $0.6949\pm 0.0082$ &  & $0.7058 \pm 0.0065$ & $0.0109$  \\
\hline 
\hline 
\end{tabular}\label{tab:besth}
\end{table}

\section{Conclusions}
\label{sec:con}

Although the concordance model has been successful in describing cosmological observations,but we can't justify the mismatch between several early and late-time inferences of the Hubble constant $H_0$ using $\Lambda$CDM, which called Hubble tension.
Depending on the dataset considered for local measurements, the significance of this tension falls between $4-6\sigma$.
In order to solve this tension, as well as the other challenge against $\Lambda$CDM, in this paper we focused on oscillating DE parameterizations. Applying the Metropolis algorithm of MCMC and using the latest observations include SnIa, BAO, CCH and CMB, we have constrained our DE models. To compare the ability of models in fitting observations, we have used AIC as a simple information criteria and DIC as a generalized form of AIC using the Kullback–Leibler divergence, which has been developed based on Bayesian statistic \citep{spiegelhalter2002bayesian,Mootoovaloo_2020,Salas_2022}. Although, DIC is really suitable for prediction, but it
has also been used for selecting the correct model.
Upon AIC results, $\Lambda$CDM selected as the best model while because of $\Delta$AIC$ >5$ with respect to $\Lambda$CDM, we could not observe any significant support for oscillating dark energy models. 
These models have $2-3$ more extra parameters with respect to concordance model which led to an increment in those AIC value. 
Upon DIC results, we observed good performance for oscillating models. It is easy to see in Tab.\ref{tab:best} that Model (5) with $\Delta$DIC$=-0.7$ is the best model which followed by $\Lambda$CDM. Moreover, we observed $\Delta$DIC$ <2$ for models (1), (4), and (6) which means "Significant support" for these models. For models (2), (7) and (8) we have "Less support" and finally "Considerably less support" for Model (3). Upon these results we conclude that some of oscillating DE models are consistent with data as well as concordance $\Lambda$ cosmology.
Furthermore, we examine the ability of oscillating models in solving the problems against $\Lambda$CDM. In the case of CCP, as one can see in Fig.\ref{fig:w}, because of oscillating behavior of these models at early time, there's no need to fine tunings and CCP will resolve. In the case of Hubble tension, comparing with $\Lambda$CDM, our models can reduce the tension (see Fig.\ref{fig:shoe}).
Comparing with $\Lambda$ based Planck measurements, our oscillating models led to a reduction of Hubble tension from 4.1-$\sigma$ to 2.14-2.56-$\sigma$ with respect to SHoES results. On the other hand, using $\Lambda$CDM in our analysis we found a 2.51-$\sigma$ tension with SHoES results which indicate that 6 models of our 8 models have better advantages with respect to concordance model in alleviating Hubble tension.
In order to find the effect of adding SHoES measurements on the results of our analysis, we repeated the analysis after adding $H_0=73.0 \pm 1.4$ from SHoES \citep{Riess:2020fzl} to our datasets. Assuming this data  point led to an increase in the best value of parameter $h$ for all of the models. These increases were relatively significant, averaging $\bar{\Delta h}=0.0095$ occurring mostly in the Model(7), Model(2) and $\Lambda$CDM (see Tab.\ref{tab:besth} and the lower panel of Fig.\ref{fig:shoe}). These results indicate that, independent of the DE models under study, assuming a single data point from SHoES measurements can change the $H_0$ results significantly. we can not relate these extra changes in $H_0$ to the ability of models in solving Hubble tension. This is in full agreement with the results of \citep{Rezaei:2023xkj}.
Summarizing the results, we can conclude that oscillating DE scenario can be assumed as an alternative for $\Lambda$CDM model. Although couldn't fit the observations as well as $\Lambda$CDM, it can solve the coincidence problem and also reduces the Hubble tension. These abilities also can be investigated in future studies using other observations at perturbation level, such as LSS data. 

\bibliography{sample631}{}
\bibliographystyle{aasjournal}



\end{document}